\documentclass[12pt,preprint]{aastex}
\slugcomment{ApJ}
\shorttitle{FORMATION OF HOT PLANETS}
\shortauthors{NAGASAWA, IDA, \& BESSHO}

\begin{document}
\title{Formation of Hot Planets by a combination of planet scattering, tidal circularization, and Kozai mechanism}
\author{M. Nagasawa, S. Ida, and T. Bessho} 
\affil{Tokyo Institute of Technology, 
2-12-1, Ookayama, Meguro-ku, Tokyo 152-8550, Japan}

\email{nagasawa.m.ad@m.titech.ac.jp}

\begin{abstract}
We have investigated the formation of close-in extrasolar 
giant planets through a coupling effect of mutual scattering, 
Kozai mechanism, and tidal circularization, by orbital integrations.
Close-in gas giants would have been originally formed at several AU's 
beyond the ice lines in protoplanetary disks and migrated close to their host stars.
Although type II migration due to planet-disk interactions may be a major
channel for the migration, we show that this scattering process
would also give a non-negligible contribution.
We have carried out orbital integrations of three planets with Jupiter-mass, 
directly including the effect of tidal circularization.
We have found that in about 30\% runs close-in planets are formed,
which is much higher than suggested by previous studies.
Three-planet orbit crossing usually results in one or two planets ejection. 
The tidal circularization often occurs during 
the three-planet orbit crossing, but previous studies have monitored 
only the final stage after the ejection, significantly underestimating
the formation probability.
We have found that Kozai mechanism by outer planets is responsible for
the formation of close-in planets.
During the three-planet orbital crossing, the Kozai excitation is repeated
and the eccentricity is often increased secularly to values close enough to unity for
tidal circularization to transform the inner planet to a close-in planet.  
Since a moderate eccentricity can remain for the close-in planet, 
this mechanism may account for the observed close-in planets
with moderate eccentricities and without nearby secondary planets.
Since these planets also remain a broad range of orbital inclinations 
(even retrograde ones), the contribution of this process 
would be clarified by more observations of Rossiter-McLaughlin effects 
for transiting planets.
\end{abstract}

\section{INTRODUCTION}
\label{sec:intro}

More than 250 planets have been detected around both solar and non-solar type stars.
Recent development of radial velocity techniques and accumulation of observations have revealed detailed orbital distribution of close-in planets. 
Figure \ref{fig:exso}$a$ shows the distribution of semi-major axis and eccentricity for 236 planets \footnote{http://exoplanet.eu/catalog-RV.php} discovered around solar type stars by the radial velocity techniques. 
The dotted line shows a pericenter distance $q=0.05$ AU. 
At $q \la 0.05$ AU, many close-in planets have small eccentricities, which are accounted for by the circularization due to the tidal dissipation of energy within the planetary envelopes (Rasio \& Ford 1996).
The semi-major axis distribution of discovered extrasolar planets shows double peaks around 0.05 AU and 1 AU (e.g., Marcy et al. 2005; Jones et al. 2005). 
The location of the outer peak is determined by the observational limits, but the inner peak at around 0.05 AU, in other words shortage of planets between 0.06--0.8 AU, would be real. 
Since the close-in planets have relatively small eccentricities, the peak around 0.05 AU appears as a pile up at $\log q/(1{\rm AU})=-1.3$ also in pericenter distribution (Fig.~\ref{fig:exso}$b$).
The close-in planets are also found in the multi-planetary systems. 
Up to now, 23 multi-planet systems are reported with known planetary eccentricities (Fig.~\ref{fig:multi}). 

The discovered close-in planets would have been formed at large distances beyond the ice line and migrated to shorter-period orbits (e.g., Ida \& Lin 2004a, 2004b). 
A promising mechanism for the orbital migration is ``type II'' migration (e.g., Lin, Bodenheimer, \& Richardson 1996). This is the migration with disk accretion due to gap opening around the planetary orbit caused by gravitational interaction between the planet and the protoplanetary gas disk. 
This mechanism can account for not only the pile-up of the close-in planets but also planetary pairs locked in mean-motion resonances.  
The migrating planets may have stalled as they enter the magnetospheric cavity in their nascent disks or interact tidally with their host stars. 
However, several close-in ($a \la 0.1$AU) planets 
such as HAT-P-2, HD118203b, and HD162020b have moderate eccentricities ($\ga 0.3$),
but are not accompanied by nearby secondary large planets. 
HD17156b with $a \simeq 0.15$AU may also belong to this group, 
since it has very large eccentricity ($\simeq 0.67$) and $q$ as small as 0.05AU.
It may be difficult for the planet-disk interaction 
alone to excite the eccentricities up to these levels.
One possible mechanism for the excitation to the moderate eccentricity for close-in planets in multiple-planet systems is the passage of secular resonances during the epoch of disk depletion (Nagasawa, Tanaka, \& Ida 2000; Nagasawa, Lin, \& Ida 2003; Nagasawa \& Lin 2005). 
Nagasawa and Lin (2005) showed that the relativity effect and the gravity of (distant) secondary planets excite the eccentricity of close-in planet orbiting around a slowly rotating host star.

Although a main channel to form close-in planets may be type II migration, inward scatterings by other giant planets can be another channel, in particular for the close-in planets with moderate eccentricities.
In relatively massive and/or metal-rich disks, multiple gas giants are likely to form (e.g., Kokubo \& Ida 2002; Ida \& Lin 2004a, 2004b, 2008).  
On longer timescales than formation timescales, the multiple gas giant systems can be orbitally unstable (Lin \& Ida 1997; Marzari \& Weidenschilling 2002). 
In many cases, a result of orbital crossing is that one of the planets is ejected and the other planets remain in well separated eccentric orbits (``Jumping Jupiters'' model; e.g., Rasio \& Ford 1996; Weidenschilling \& Marzari 1996; Lin \& Ida 1997).
Recent studies of the Jumping Jupiter model show pretty good coincidence with the observational eccentricity distribution (Marzari \& Weidenschilling 2002; Chatterjee, Ford, \& Rasio 2007;  Ford \& Rasio 2007). 

With the sufficiently small pericenter distance ($\la 0.05$ AU), the planetary orbit 
can be circularized into the close-in orbit by tidal effects from the star (Rasio \& Ford 1996). 
In this paper, the process of planet-planet scattering followed by the tidal
circularization will be referred to as "scattering model."
To send a planet into the inner orbit, the other planets have to lose their orbital energy. 
Since inward scattering of a lighter planet is more energy saving, 
the systems which experienced the planet-planet scattering tend to have smaller planets inside. 
We might see the tendency in Figure \ref{fig:multi}, but it is also true that radial velocity technique tends to detect heavier planets in outer regions. 
Note that since outer planets have negligible orbital energy, the energy conservation cannot restrict the semi-major axes of the outer planets and the outer planets may be located 
to the regions far beyond the radial velocity technique limit (Marzari \& Weidenschilling 2002).

Observation of Rossiter-McLaughlin effect for 5 transiting close-in planets 
including HAT-P-2 with eccentricity $e \simeq 0.52$
show that their orbital planes are almost aligned with the stellar spin axes (Narita et al. 2007; Winn et al. 2005, 2006, 2007; Wolf et al. 2007), which may suggest that the scattering 
mechanism is not a main channel for formation of close-in planets, 
because the close scattering may excite orbital inclination as well as eccentricity 
(note that as shown in \S 3.4, the orbital inclinations of close-in planets formed by 
the scattering model are not necessarily high).
Since HD17156b with $e \simeq 0.67$ is also transiting,
the observation of Rossiter-McLaughlin effect will be important.

More serious problem of the scattering model 
may be the probability for pericenter distances to become small enough (in other words, for eccentricities to become close enough to unity (e.g., $e \ga 0.98$)) to allow tidal circularization.
In a system with only two giant planets, the energy conservation law keeps the ratio of semi-major axes close to the initial value (see \S \ref{subsec:ppscat}). 
In such system, the probability for final planets with a large ratio of semi-major axes and very large eccentricity is $\la 3$\% (Ford, Havlickova, \& Rasio 2001). 
In a system with more giant planets, the situation is slightly improved, 
especially when planets have non-zero inclinations. 
The close-in planets are very rare just after the scattering, but when they take into account the longer orbital evolution, the probability of the candidates for close-in planets increases to $\sim$10 \% (Weidenschilling \& Marzari 1996; Marzari \& Weidenschilling 2002; Chatterjee et al. 2007). 
The contributed planets are planets in Kozai state (see \S 3.5).

Here we revisit the scattering model through orbital integration of three giant planets with direct inclusion of the tidal circularization effects and analytical argument of Kozai effect. 
We find that the probability for formation of close-in planets is remarkably increased to $\sim$30\%. 
Although the previous studies were concerned with orbital states after stabilization by ejection of some planet(s), we find that the orbital circularization occurs through ``Kozai migration'' (Wu 2003; Wu \& Murray 2003) caused by outer planets, mostly during three-planet orbital crossing, after a tentative separation of the inner planet and the outer planets.

In the next section, we describe orbital instability of a planetary system (\S \ref{subsec:ppscat}) and the orbital evolution that becomes important after the system enters stable state: tidal circularization (\S \ref{subsec:tide}) and Kozai mechanism (\S \ref{subsec:kozai}). 
Following description of its models and assumptions, we present results of numerical simulations in \S \ref{sec:numerical}. 
We show typical outcome of the scattering (\S \ref{subsec:resultscat}), how the planets are circularized into short-period orbits (\S \ref{subsec:resulteach}), and its probability (\S \ref{subsec:resultcir}). 
In \S \ref{sec:analy} we present analytical arguments. 
We summarize our results in \S \ref{sec:conclusion}.

\section{THE PLANET-PLANET SCATTERING AND THE TIDAL CIRCULARIZATION}
\label{sec:basic}

In this paper, we carry out orbital integration of three equal-mass ($m$) giant planets with physical radii $R$ rotating around a host star with mass $m_{\ast}$. 
Interaction with disk gas is neglected. 
We consider the case in which the timescale of orbital crossing to occur is longer than disk lifetime (see \S 2.1).
The orbits are integrated until one planets is tidally circularized, two planets are ejected, or the system becomes stable after one planet is ejected and the others are orbitally well separated.
But, we keep integrating long enough after the orbital stabilization to see Kozai mechanism that occurs on longer timescales. 
To understand the numerical results, in this section, we briefly summarize key processes for our calculations: orbital instability, tidal circularization, and Kozai mechanism. 

In the present paper, Jupiter-mass planets ($m=m_{\rm J}=10^{-3}M_{\odot}$) and a solar-mass host star ($m_{\ast}=M_{\odot}$) are adopted.
In the following, we use $a$ for semi-major axis, $e$ for eccentricity, $i$ for inclination, and $\omega$ for argument of pericenter. 
When we distinguish the planets, we use subscripts 1, 2, and 3 from inner to outer planets.
A subscript '$\ast$' shows quantities for the host star and a subscript 'tide' is used for values related to dynamic tide.
A variable $h$ is defined as $h= (1-e^2)\cos^2i$. 
Mutual Hill radius of planets $k$ and $j$ is expressed as 
$R_{\rm H}(k,j)= [(m_k+m_j)/3m_{\ast}]^{1/3}(a_k+a_j)/2$.

\subsection{Planet-Planet Interactions}\label{subsec:ppscat}

Timescales for orbital instability to occur ($T_{\rm cross}$) very sensitively depends on orbital separation between planets, their eccentricities, and masses (Chambers, Wetherill, \& Boss 1996; Chambers, \& Wetherill 1998; Ito and Tanikawa 1999, 2001; Yoshinaga, Kokubo, \& Makino 1999).
Chambers et al. (1996) found that for systems with equal-mass ($m$) planets in initially almost circular orbits of orbital separation $b R_{H}$, 
$\log T_{\rm cross} = \alpha b +\beta$, where $\alpha$ and $\beta$ are constants in terms of $b$.
The values $\alpha$ and $\beta$ weakly depend on $m$ and number of planets.
Marzari and Weidenschilling (2002) showed that for three Jupiter-mass planets with
$a_1=$5 AU, $a_2=a_1+b R_{\rm H}(1,2)$, $a_3=a_2+b R_{\rm H}(2,3)$, 
$\alpha \simeq 2.47$, and $\beta \simeq -4.62$.
Chatterjee et al. (2007) derived an empirical formula that is not skewed towards greater timescales: The median of $\log T_{\rm cross}$ is given by 
$1.07 + 0.03 \exp{(1.10 b)}$. 
Since the resonances modify $T_{\rm cross}$, especially in systems with massive planets, a real timescale is sometimes much shorter than above timescale. 
But these expressions are helpful to know how long we need to continue the simulations.

Note that semi-major axis of an final innermost planet ($a_{\rm fin,1}$) is limited by energy conservation when only mutual scattering is considered.  
In an extreme case, in which the other planets are sufficiently far away 
($a_{{\rm fin},k} \gg a_{{\rm fin},1}$; $k=2,3,...$),
\begin{equation}
\frac{1}{a_{\rm fin,1}} \simeq \sum_k^N \frac{1}{a_{{\rm ini}, k}},
\label{eq:egconserve}\end{equation} 
where $a_{{\rm ini}, k}$ is the initial semi-major axis of planet '$k$' 
and $N$ is  number of planets. 
The equation implies that $a_{\rm fin,1}$ should be larger than $a_{{\rm ini},1}/N$.
That means close-in planets are hardly formed as a result of direct scattering of a few planets that are originated from the regions beyond several AU like Jupiter and Saturn. 

\subsection{Tidal Circularization}\label{subsec:tide}

As shown above, the planet-planet scattering cannot make the orbital period of 
a planet as short as a few days, as long as the energy is conserved. 
However, if energy is dissipated in the planetary interior by tidal force from the host star, a planet with a few day period can be formed. 
With $e$ highly excited to a value close to unity, the pericenter of the planet can approach to its host star, since pericenter distance is given by $q=a(1-e)$.
Then, both $e$ and $a$ are decreased with keeping $q$ almost constant by the tidal dissipation.

The tidal dissipation is a strong function of $q$.
As the scattered planets normally have highly eccentric orbits, we only consider dynamic tide. 
In our simulation, we adopt the formula by Ivanov and Papaloizou (2004).
They analytically calculated the strongest normal modes, 
$l = 2$ fundamental modes, of the tidal dissipation.
It is assumed the normal modes arisen near the pericenter are fully dissipated before a next pericenter passage. 
If the normal modes remain, the angular momentum and energy can either increase or decrease at the new pericenter passage (e.g., Mardling 1995a, 1995b).

Ivanov and Papaloizou (2004) derived the tidally gained angular momentum 
($\Delta L_{\rm tide}$) and energy ($\Delta E_{\rm tide}$) during
a single pericenter passage as
\begin{eqnarray}
\Delta L_{\rm tide} \sim& &\!\!\!\!\!\!\!\!
-\frac{32 \sqrt{2}}{15}\tilde{w_0}^2\tilde{Q}^2\xi
\exp\left(-\frac{4\sqrt{2}}{3}w_+\xi \right) 
\left[1-\frac{9}{2^{14}(\tilde{w_0}\xi)^4}
\exp \left( \frac{4\sqrt{2}}{3} \tilde{\sigma}\xi \right) \right] 
L_{\rm pl}, \label{eq:ang}\\
\Delta E_{\rm tide} \sim& &\!\!\!\!\!\!\!\!
-\frac{16 \sqrt{2}}{15}\tilde{w_0}^3\tilde{Q}^2\xi
\exp\left(-\frac{4\sqrt{2}}{3}w_+\xi\right) 
\left[1+\frac{3}{2^7(\tilde{w_0}\xi)^2}
\exp\left(\frac{2\sqrt{2}}{3} \tilde{\sigma}\xi \right) \right]^2 
E_{\rm pl},
\label{eq:eng}\end{eqnarray}
where $L_{\rm pl}=m(GmR)^{1/2}$ and $E_{\rm pl}=Gm^2/R$ are
orbital angular momentum and orbital energy, 
$\xi$=$(mq^3)^{1/2}(m_{\ast}R^3)^{-1/2}$, 
$w_+\sim\tilde{w_0}(Gm/R^3)^{1/2}+\Omega_r$,
and $\tilde{\sigma}\sim 2\Omega_r/(Gm/R^3)^{1/2}$.
The value $\Omega_r$ is angular velocity of the planet rotation, $\tilde{w_0}$ is 
a dimensionless frequency of fundamental mode, and $\tilde{Q}$ is a 
dimensionless overlap integral that depends on the planetary interior model. 
The spin axis is assumed to be perpendicular to the orbital plane.
We use the same planetary model as Ivanov and Papaloizou (2004). 
From their Figure 6 and Figure 7, we approximate 
$\tilde{w_0}\simeq 0.53(R/R_{\rm J})+0.68$ and 
$\tilde{Q}\simeq -0.12(R/R_{\rm J})+0.68$ for Jovian mass planet, where 
$R_{\rm J}$ is Jovian radius. 

For a non-rotating planet, equations (\ref{eq:ang}) and (\ref{eq:eng}) are simplified to
\begin{eqnarray}
\Delta L_{\rm tide} \sim& &\!\!\!\!\!\!\!\!
-\frac{32 \sqrt{2}}{15}\tilde{w_0}^2\tilde{Q}^2\xi 
\exp\left(-\frac{4\sqrt{2}}{3}w_0\xi \right) L_{\rm pl}, 
\label{eq:lnorot}
\\
\Delta E_{\rm tide} \sim& &\!\!\!\!\!\!\!\!
-\frac{16 \sqrt{2}}{15}\tilde{w_0}^3\tilde{Q}^2\xi
\exp\left(-\frac{4\sqrt{2}}{3}w_0\xi\right) E_{\rm pl}
.
\label{eq:enorot}\end{eqnarray}
When the tide spins the planet up to its critical rotation, 
$\Omega_r = \Omega_{\rm crit}$, which gives no additional increase in planetary angular momentum in equation (\ref{eq:ang}),
\begin{eqnarray}
\Delta L_{\rm tide} =& &\!\!\!\!\!\!\!\! 0,
\label{eq:lcrit}
\\
\Delta E_{\rm tide} \sim& &\!\!\!\!\!\!\!\!
-\frac{1}{5\sqrt{2}}\frac{\tilde{w_0}\tilde{Q}^2}{\xi}
\exp\left(-\frac{4\sqrt{2}}{3}w_0\xi\right) E_{\rm pl}
.
\label{eq:ecrit}\end{eqnarray}
The values of $\Delta E_{\rm tide}/E_{\rm pl}$ for a $m_{\rm J}$ planet 
around a $M_{\odot}$ star given in equations (\ref{eq:enorot}) and 
(\ref{eq:ecrit}) are shown in Figure \ref{fig:deleng} for 
$R=R_{\rm J}$ and $R=2R_{\rm J}$ ($m=m_{\rm J}$).
As the figure shows, the tide is only effective in the vicinity of the star. 

The model is only valid for fully convective planets in highly eccentric orbits, since it considers only $\ell=2$ f-mode and uses impulse approximation. 
When a planetary orbit is considerably circularized to a level that the impulse approximation becomes invalid and a quasi-static tide becomes important, the above equations are no longer relevant enough. 
However, since our purpose of this paper is not to study the details of tidal circularization process itself but to show an available pass to formation of close-in planets, we use the above formulae for dynamic tide until the end of simulations.

With estimation
$d E_{\rm tide}/dt \simeq \Delta E_{\rm tide}/T_{\rm Kep}$ and  
$d L_{\rm tide}/dt \simeq \Delta L_{\rm tide}/T_{\rm Kep}$
(the pericenter passage occurs every $T_{\rm Kep}$), the evolution timescale of the semi-major axis and eccentricity are 
\begin{eqnarray}
\tau_a&=&-\frac{a}{\dot{a}}=\frac{Gmm_{\ast}}{2a}
\frac{T_{\rm Kep}}{(-\Delta E_{\rm tide})}, \label{eq:a_damp_timescale}\\
\tau_e&=&-\frac{e}{\dot{e}}=Gmm_{\ast}T_{\rm Kep}
\left[-a\gamma\Delta E_{\rm tide}+\sqrt{\frac{Gm_{\ast} \gamma}{a e^2}} 
\Delta L_{\rm tide}\right]^{-1},
\label{eq:e_damp_timescale}\end{eqnarray}
where $T_{\rm Kep}$ is a Keplerian time and $\gamma=(1-e^2)/e^2=q(2a-q)/(a-q)^2$.
The damping timescale for a planet scattered into 
a highly eccentric ($e \sim 1$) orbit at $a = 2$AU 
is shown in Figure \ref{fig:timescale} ($q = a(1-e)$). 
The stellar and planet masses are 1$M_{\odot}$ and $m_{\rm J}$, respectively. 
Two limiting cases with strong tide ($\Omega_r=0$ and $R=2 R_{\rm J}$) and weak tide 
($\Omega_r=\Omega_{\rm cirt}$ and $R=1 R_{\rm J}$) are plotted. 
In the strong tide case, the semi-major axis is damped within $10^{10}$ years at 
$q \la 0.04$ AU. 
In the weak damping case, $q \la 0.02$ AU. 
In the following numerical simulations we test both cases.

\subsection{Kozai Mechanism}\label{subsec:kozai}

How closely the pericenter approaches the host star depends on eccentricity.
Even after a planet is settled in a stable orbit, its eccentricity and inclination can largely oscillate by the secular perturbation from separated other planets, in particular, when the inner planet has relatively small $z$-component of the angular momentum ($\ell_z$). 
The mechanism is called Kozai mechanism (Kozai 1962). 
Suppose that the inner planet '1' is perturbed by a well-separated outer planet '2'.
Eccentricity $e_1$, inclination $i_1$, and argument of pericenter of the inner planet evolve conserving its time averaged Hamiltonian and $\ell_z$.
In the lowest order of $a_1/a_2$ ($a_1$ and $a_2$ are conserved), the conservation of the time averaged Hamiltonian of the inner planet and that of $\ell_z$ give
\begin{eqnarray}
(2+3e_1^2)\left(\frac{3h}{1-e_1^2}-1\right) \!\!&+&\!\! 15e_1^2
\left(1-\frac{h}{1-e_1^2}\right)\cos(2\eta) \equiv  C = {\rm const}.,
\label{eq:hamilton} \\
(1-e_1^2)\cos^2i_1 \!\!&\equiv&\!\! h = {\rm const}., 
\end{eqnarray} 
where $\eta$ is difference of argument of pericenter of the two planets 
(i.e., $\eta=\omega_2-\omega_1$).
The above equations are independent of $a_1$ and parameters of the outer planet, but the timescale for the circulation or the libration is in proportion to 
$(1-e_2)^{3/2}(a_2/a_1)^3 m_2^{-1} n_2^{-1}$,
where $n_2$ is mean motion of planet 2. 
The conservation of $\ell_z$, i.e., $h={\rm const.}$, shows that change in $i_1$ 
toward 0 results in increase in $e_1$ toward unity, 
i.e., approach of the pericenter to the vicinity of the star.
How closely $e_1$ can approach unity is regulated by eq.~(\ref{eq:hamilton}).

\section{THE NUMERICAL SIMULATIONS}
\label{sec:numerical}

\subsection{Initial Setup}\label{subsec:initial}

We study evolution of systems of three planets with Jupiter-mass 
($m = 10^{-3}M_{\odot}$) orbiting a solar mass star in circular orbits 
($e_1=e_2=e_3=0$). 
Their initial semi-major axes are $a_1=5.00$ AU, $a_2=7.25$ AU, 
and $a_3=9.50$ AU and their inclinations are $i_1=0.5^{\arcdeg}$, 
$i_2=1.0^{\arcdeg}$, and $i_3=1.5^{\arcdeg}$,
following the simulation setup of Marzari and Weidenschilling (2002). 
Orbital angles are selected randomly.
We repeat the orbital integration with different seeds of random number generation for the initial orbital angles with the same initial $a$, $e$ and $i$.

With this choice, the shortest semi-major axis after the scattering is expected to be $a_{\rm min}=2.26$ AU (eq.~[\ref{eq:egconserve}]). 
The tidal damping timescale is a function of mass and radius of planets. 
Although we fix the planetary mass, we test $R=R_{\rm J}$ and $2 R_{\rm J}$ cases. 
The latter case corresponds to newborn planets that have not been cooled down.
Since an averaged orbital separation of the system is $\sim 3.6 R_{\rm H}$ in this choice of semi-major axes, it is expected that the orbital instability starts in timescales of $\sim 10^3$ years (see \S \ref{subsec:ppscat}). 

Oscillation modes are raised in the planetary interior by the tidal force from the host star in the vicinity of the pericenter. 
We assume that the energy of the modes is dissipated and the angular momentum is transferred to the orbital angular momentum before the next pericenter passage. 
Assuming that the orbital changes are small in individual approaches, we change the orbit impulsively at the pericenter passage as mention below. 

We have performed 6 sets of simulations (Table \ref{tab:set}). 
In Set V, we adopt a simplest model, that is, the velocity (${\bf v}$) of the planet is changed discontinuously to ${\bf v}'$  at the pericenter passage as
\begin{equation}
{\bf v}^{\prime}= \sqrt{2\Delta E_{\rm tide} +v^2}\frac{{\bf v}}{v},
\label{eq:vchange}
\end{equation} 
where $\Delta E_{\rm tide}$ is given by equation (\ref{eq:enorot}).
In this set, we adopt $\Omega_r=0$ and $R=2 R_{\rm J}$. 
Since we do not change the location of the pericenter and direction of motion there, the angular momentum variation is specified as $q\Delta v$, which is inconsistent with equation (\ref{eq:lnorot}). 
In Sets T1 and T2, on the other hand, the pericenter distance is also changed as well as velocity so as to  consistently satisfy both equations (\ref{eq:lnorot}) and (\ref{eq:enorot}) for $\Omega_r = 0$, but the direction of the motion does not change. 
Equations (\ref{eq:lcrit}) and (\ref{eq:ecrit}) are satisfied in Set T3 and T4 ($\Omega_r = \Omega_{\rm crit}$).
Planetary radius is $R=2 R_{\rm J}$ in Sets T1 and T3, while $R = R_{\rm J}$ in Sets T2 and T4. 
As we will show later, the probability for the tidal circularization to occur does not significantly depend on the models of orbital change by tidal dissipation. 
For comparison, we also carry out the case without the tidal circularization (Set N).

Because of the chaotic nature of the scattering processes, we integrate 100 runs in each set. 
We integrate orbits for $10^7-10^8$ years.
We stop the calculation when a planet hits the surface of the host star with 1 Solar Radius or when $\Delta L_{\rm tide}$ overcomes the angular momentum that a circularized planet has. 
The latter condition happens in Sets T1-T4. 
In Sets V, a collision against the host star occurs to a tidally circularized planet. 
In other cases, we check the stability of the systems in every $10^6$ years after $10^7$ years.
If only one planet survives and its pericenter is far from the star, or two are left with dynamically stable orbits with relatively large $\ell_{z}$, we stop the simulation. 
We continue the simulation until $10^8$ years as long as the system contains three planets.

\subsection{Outcome of Planet-Planet Scattering: Set N}
\label{subsec:resultscat}

The result of the planet-planet scatterings without tidal force is consistent with previous studies of Marzari and Weidenschilling (2002). 
The systems that ended with two planets are the most common outcome. 
In 75 cases of Set N, one planet is ejected.
In 22 cases, one of the planets hits the host star.
It mainly occurs during chaotic phase of planetary interaction, namely, before first ejection of a planet. 
The case in which 
two planets are ejected is rare as Marzari and Weidenschilling found. 
We observed such outcome in 5 runs. 

The distribution of semi-major axis and eccentricity of the final systems is shown in Figure \ref{fig:nonea}.  
The innermost planets are scattered into orbits at $a \simeq 2.5$ AU. 
Since a small difference in the orbital energy causes large difference in the semi-major axis in the outer region, the semi-major axes of outer planets are widely distributed. 
The figure includes planets that hit the star. 
They are clumped around $e \simeq 1$.
The planets with small pericenter distance $q \la 0.05$ AU are 
composed of the star-colliding planets. 
Since there is no damping mechanism in Set N, 
the small $q$ planets also have $a \sim 2$--3 AU.

\subsection{Orbital Evolution to Hot Planets}
\label{subsec:resulteach}

The star-approaching planets are circularized to 
become close-in planets when we include tidal force in our simulation. 
Typical evolution of semi-major axis, pericenter, and apocenter in the case of Set V is shown in Figure \ref{fig:ex1}. 
The system enters chaotic phase quickly and originally the outermost planet is 
scattered inward into $a \simeq 3$ AU through several encounters. 
The planet is detached from other planets and becomes marginally stable with 
$a \sim 3$ AU after $t \sim 10^5$ years until it suffers tidal damping 
($t \ga 3.9 \times 10^6$ years), though the outer two planets still continue orbital crossing. 
During the tentative isolation period, the eccentricity and inclination of the innermost planet are mostly varied by secular (distant) perturbations from the outer two planets (Figure \ref{fig:ex1ei}). 
Since the perturbations are almost secular, the semi-major axis of the isolated planet does not change largely until $3.9 \times 10^6$ years, but its eccentricity randomly varies at the occasions when the middle planet approaches. 
As a result of one of these repeated close encounters, the isolated planet acquires relatively large $e$ and $i$ at $t \simeq 1.0 \times 10^6$ years. 
Its eccentricity oscillates with large amplitude, exchanging $\ell_z$ with the inclination by Kozai mechanism.
Although the amplitude of oscillations of $e$ and $i$ are decayed after 
$t \simeq 1.7 \times 10^6$ years, they are pumped up again at 
$t \simeq 2.5 \times 10^6$ years.  
At $3.7 \times 10^6$ years, the planet acquires very large oscillation amplitude of inclination and eccentricity.
The eccentricity reaches the maximum value of a Kozai cycle at $3.9 \times 10^6$ years, 
at which the pericenter approaches to the host star. 
Then the planet is tidally moved slightly inward, but the damping of the semi-major axis is interrupted when the eccentricity reaches the maximum value of a Kozai cycle and turns to decrease. 
At $4.1 \times 10^6$ years, the pericenter distance $q$ can be small enough for tidal circularization during Kozai cycle. 
Since $q$ becomes $<0.01$ AU in this case, the orbit is circularized on timescale of $10^4$ years.

Another example from Set T1 is shown in Figure \ref{fig:ex2}. 
In this case, one of the planets enters a hyperbolic orbit at 2.2 $\times 10^6$ years. 
Two planets are left in stable orbits at 2.3 and 80 AU. 
However, the innermost planet is in the Kozai state. 
At the time of isolation, the eccentricity and inclination of the innermost planet are 0.49 and 1.5 radian, respectively. 
The eccentricity and inclination oscillate by the Kozai mechanism. 
Because the perturbing planet is located much further ($\sim 80$ AU) than in the previous case in Figure \ref{fig:ex1}, eccentricity increases more slowly. 
At 10.5$\times 10^6$ years, the eccentricity reaches 0.984.
Since $q$ reaches $\simeq 0.02$ AU, the planet's $e$ and $a$ are tidally damped.
Since the damping timescale of the eccentricity ($\tau_e$) is longer than that of the semi-major axis (see Fig.~\ref{fig:timescale}), the eccentricity decays after the semi-major axis significantly decreases.  
In contract to Set V, $q$ gradually increases during the tidal circularization in Set T. 
Since the damping timescale is a strongly increasing function of $q$,
the tidal circularization slows down with time in this Set. 
As a result, the eccentricity is not fully damped in $10^8$ years
($e=0.08$ at $10^8$ years in this example).
Wu (2003) and Wu and Murray (2003) also pointed out the migration due to 
a coupled effect of Kozai mechanism and tidal circularization 
for a planet in a binary system and called ``Kozai migration.'' 
They considered Kozai mechanism induced by perturbations of a companion star,
while we consider that of by outer planets in orbital crossing.
If the planet scattered inward has small $h=(1-e^2)\cos^2 i$,
it is subject of the tidal circularization.

\subsection{Final Orbital Distribution}
\label{subsec:orbitdist}

The distribution of final eccentricity of innermost planets is shown in Figure \ref{fig:histe}. 
It is divided into three groups: a peak at $e>0.95$, that at $e < 0.05$, and a broad distribution between the peaks. 
The peak at $e>0.95$ is composed mainly of planets in Set N (long-dashed line). 
This reflects a fact that we have stopped the simulation in Set N when the planet hits the surface of the host stars. 
In other sets, these planets are tidally circularized. 
Almost all the circularized planets go to the most prominent peak at $e<0.05$. 
Excluding these two peaks, the eccentricity is broadly distributed centered at 
$e \sim 0.5$, as previous authors found. 

The planets that are injected to the inner orbits with moderate eccentricities but have not suffered tidal circularization are distributed at around $a\sim 2.3$ AU, as the energy conservation law requests. 
The pericenter distribution is shown in Figure \ref{fig:histq}. 
The distributions at $\log (q/1{\rm AU})>-0.5$ have no remarkable difference between models. In Set V, the close-in planets are piled up at $\log (q/1{\rm AU})=-2.3$. 
Since we do not take into account the effect
that the dynamic tide should become less efficient according to orbital circularization, the planet cannot stop the tidal decay once the circularization is started, which leads to the overdensity at the lower-$q$ peak. 
Such a lower peak does not exist in Sets T, 
since the increase in $q$ during the tidal circularization slows down
the further circularization, as shown in Figure \ref{fig:ex2}. 
In Set T3 and T4 in which $\Delta L_{\rm tide} = 0$,
the initial angular momentum, 
$\sqrt{G m_* (1-e_{\rm ini}^2)a_{\rm ini}} \simeq 
\sqrt{2G m_* q_{\rm ini}}$ is equal to the final one,
$\sqrt{G m_* (1-e_{\rm final}^2)a_{\rm final}}
\simeq \sqrt{G m_* q_{\rm final}}$, so $q$ can increase by a factor 2.
The final $q$ of close-in planets are distributed between 
$\log (q/1{\rm AU})=-2$ and $-1$ like the observed planets 
(see Fig. \ref{fig:exso}$b$).
The final $e$ are not necessarily damped fully and distributed in a range of 0--0.4 in Sets T,
and the final $a$ are larger than that in Set V.

The inclination distribution is shown in Figure \ref{fig:histi} in the cases of Set N (panel $a$) and Set T1 (panel $b$). 
Solid lines show the distribution of only innermost planets. 
Dashed lines show that of all remaining planets (inner $+$ outer planets).
The overall distributions show no difference between models with and without tide. 
Most planets keep relatively small inclinations as Chatterjee et al. (2007) claimed.
However, the close-in planets shown in panel $c$ tend to have relatively large inclination, because the Kozai migration is effective for planets injected to highly inclined ($i \sim \pi/2$) orbits so that close-in planets formed with this model selectively have highly inclined orbits.
In the Kozai mechanism, however, the inclination takes lower values 
when the eccentricity is higher. The tidal circularization occurs when the eccentricity is high.
Thus the close-in planet is formed more easily when the inclination takes lower values in Kozai cycle. 
This effect inhibits over-density at the highest inclination and results in rather broad inclination distribution.
Note that we did not include any damping mechanisms for inclination, which generally have very long timescales.
Our model shows that non-negligible fraction of close-in planets have retrograde rotation ($i > \pi/2$) with respect to the equator of the host star, 
although the retrograde planets can be tidally unstable on longer timescales.
Since type II migration cannot produce retrograde rotation, if such 
retrograde planets are discovered, it would be a proof of some contribution of
the scattering model to formation of close-in planets.

\subsection{Circularization Probability}
\label{subsec:resultcir}

The fraction of runs in which close-in planets are formed is shown in Table \ref{tab:set} for each model.
It is $\sim$30\% almost independent of the detailed model for tidal circularization.
This is much higher than the probability ($\sim 10$\%) of formation of close-in planet candidates predicted by Marzari and Weidenschilling (2002).
We found that in $\sim 2/3$ of runs that form close-in planets, the planets are produced during three-planet interactions. 
In the residual $\sim 1/3$ runs, the close-in planets are formed after one of the planet is ejected from the system and the system becomes stable. 
The difference between our result and the prediction by Marzari and Weidenschilling (2002) comes from the fact that Marzari and Weidenschilling (2002) were concerned only with the latter cases after stabilization and did not pay attention to the former cases that contribute much more.
The latter circularization cases is 6--8\% in total simulations, which is consistent with the estimation of Marzari and Weidenschilling (2002).

Tidal circularization models we used could be too strong, since we use the formula of dynamic tide even after the eccentricity is significantly damped. 
However, even if a more realistic model is used, the result would not change significantly.
We used 5 different tidal circularization models.
Among our models, the damping force is the weakest in Set T4 
(see Fig.~\ref{fig:deleng}). 
In this case, only planets with $q \la 0.02$ AU are circularized. 
Nevertheless, the probability for formation of close-in planets is still as high as $\sim 30$\%. 
The relatively high probability is resulted by repeated Kozai mechanisms that are inherent in the scattering of three giant planets.

Figure \ref{fig:nonlong} shows an example of $e$, $i$, and $h$ evolution without tidal force. 
The evolution of semi-major axis of this example is shown in Figure \ref{fig:nonlonga}. 
During the initial phase ($t < 1.5 \times 10^6$ years) in which all the planets repeat mutual close encounters, their eccentricity is chaotically changed. 
However, after a planet is scattered into an inner orbit of $a = 2.47$ AU, its orbital change is mostly regulated by secular variation.  
Anti-correlation between the magnitude of $e$ and that of $i$ shows that the planet is in Kozai state. 
As we will explain in the next section, the maximum eccentricity during Kozai cycle is determined mainly by value of $h$, almost independent of the location and mass of the perturbing planet.  
Since there is no energy dissipation in this run (even with tidal circularization, energy dissipation is practically negligible as long as $q \ga 0.05$ AU), the isolation of the innermost planet is tentative and it occasionally undergoes a relatively close encounter with the outer planet(s).
After the encounter, the planet enters new Kozai state with a changed $h$. 
In the case of Figure \ref{fig:nonlong}, the planet enters Kozai state at 
$t \simeq 1.5 \times 10^6$ years first. 
This Kozai cycle is terminated by a relatively close encounter with an outer planet before eccentricity is fully increased.
The next Kozai state starts at $t \simeq 2.5\times 10^6$ years, but $h$ of this cycle is not small enough to raise $e$ to $\ga 0.98$ ($q \la 0.05$ AU). 
The third Kozai state starting at $t \simeq 4.5 \times 10^6$ years is not clean Kozai state because an outer planet is relatively close.
However, thanks to the small value of $h$,  $e$ can be $\ga 0.99$ ($q \la 0.02$ AU) in the cycle. 
If the tidal damping is included, the orbit would be circularized.
Eventually, at $t \simeq 6.5 \times 10^6$ years, one of the outer planets is ejected and the system enters a stable state where the inner planet and the remaining outer planet are separated by 70 AU. 
At the point when the system enters the stable state, $q \sim 0.4$ AU.
However, since the inner planet is again in Kozai state with relatively small $h$, the pericenter periodically approaches 0.04 AU, where the tidal dissipation is marginally effective in a relatively strong tide model. 
Since the outer planet is separated by 70 AU, the period of Kozai cycle is more than $10^7$ years.

Marzari and Weidenschilling (2002) noticed this long term orbital change after the stabilization that can bring the pericenter to the vicinity of the stellar surface, although they did not refer to Kozai mechanism.
But, with their criterion, the third Kozai cycle in which $q$ takes the smallest value in this run is missed.  
During three-planet orbit crossings, there are many chances to be in Kozai state.
And the Kozai state is quasi-stable and it often has enough time for the pericenter to stay close to the star and be tidally circularized.
Since we monitored $q$ also during the three-planet orbit crossing (in the models with tidal circularization effect, it is automatically monitored), we found much more cases to form close-in planets than previous authors did.

Figure \ref{fig:oscie} gives another example. 
In this example, dynamical tide is included. 
After the chaotic phase ends, three planets enter a quasi-stable state. 
The eccentricity of innermost planet oscillates with large amplitude exchanging $e$ and $i$. 
The longer period perturbation gradually enhances the mean eccentricity, and the planet's orbit is circularized at $1.6 \times 10^7$ years. 
This is a similar example that Marzari and Weidenschilling (2002) found. 
The close-in planet can be formed even if it is not injected into a good Kozai condition at once as long as the planetary interaction continues.

\section{THE PATH TO KOZAI MIGRATION}
\label{sec:analy}

In previous section, we have showed that the Kozai mechanism coupled with tidal circularization effectively sends planets to short-period orbits. 
In this section, we will present analytical arguments on Kozai mechanism to evaluate the probability of forming close-in planets, to explain the numerical results.

In Figure \ref{fig:hamap}, contours of $C$ given by equation (\ref{eq:hamilton})  (``Hamiltonian map'') is plotted on the $e-\eta$ plane, where 
$\eta = \omega_2-\omega_1$.
The topology of the Hamiltonian map depends only on $h$.  
The upper and lower panels are the maps with $h=0.2$ and $h=0.6$, respectively. 
When a planet is scattered into an orbit having some $(C, h)$, the planet's orbit 
moves along the constant $C$ line that is determined by $h$, provided the other planets are well separated and perturbations from them are secular. 

The argument of pericenter ($\eta$) liberates when $C<6-h$ and $h<0.6$. 
With other sets of $(C,h)$, $\eta$ circulates. 
The range of $C$ and $\sqrt{h}$ that the planet can actually has is determined by conditions of $0\le e <1$,  $0\le i \le \pi$, and $0\le \eta \le 2\pi$.
In Figure \ref{fig:chrange}, we plot the allowed range of $C$ and $h$.
At the libration center, $C$ takes the minimum value of $-20-24h+12\sqrt{15 h}$. Corresponding eccentricity is $(1-\sqrt{5h/3})^{1/2}$.

As the figure shows, the attainable maximum eccentricity ($e_{\rm max}$) is higher for a lower $h$ value.
The maximum eccentricity achieved at $\eta=\pi/2$ is 
\begin{equation}
e_{\rm max}^2=\frac{1}{36} 
\left(16-24h-C+\sqrt{400-1200h+40C+576h^2+48hC+C^2} \right).
\label{eq:emax}\end{equation}
For $e_{\rm max}$ closer to unity, the minimum pericenter distance during Kozai cycle is smaller.
If it is small enough, tidal circularization is efficient and formation of close-in planet is expected. 
In Figure \ref{fig:chrange}, contours of $e_{\rm max}$ for values close to unity are plotted on the $C$-$\sqrt{h}$ plane.  
Since $\sqrt{h}$ is proportional to angular momentum $\ell_z$, this plane is essentially the energy-angular momentum plane.
Assuming that $C$ and $\sqrt{h}$ is distributed homogeneously after scattering, though this assumption may not be exactly true, the probability ($P(e_{\rm crit})$) that a planet enters regions of $e_{\rm max} > e_{\rm crit}$ is 46.8\%, 30.8\%, and 22.1\% for $e_{\rm crit} = 0.95, 0.98,$ and 0.99, respectively.

In our simulation in which an initially innermost planet is at 5 AU, a typical final semi-major axis of innermost planet is $\sim 2$ AU (Figure \ref{fig:nonea}).
With the tidal circularization models we adopted, the circularization is effective on timescales $\la 10^8$ years if pericenter distance is smaller than $\sim 0.015$--0.03 AU. 
Then $e_{\rm max} \ga e_{\rm crit} = 0.985$--0.993 is required for the circularization.
From the probability $P(e_{\rm crit})$ estimated above, we expect that formation probability of close-in planets is 20--30\% for this setting, which is consistent with our numerical results.
Realistic distribution of $C$ and $\sqrt{h}$ might be more concentrated in lower $C$ and/or higher $\sqrt{h}$.
In this case, $P(e_{\rm crit})$ may be smaller for one scattering, but the repeated nature of Kozai mechanism may compensate for the decrease in $P(e_{\rm crit})$ to have the formation probability of close-in planets be a similar level.
Our numerical simulations in \S\ref{sec:numerical} were done for only one initial setting of the planetary semi-major axis. But the above estimates suggest the formation probability of the close-in planets may be similar in other settings.

\section{CONCLUSION AND DISCUSSION}
\label{sec:conclusion}

``Standard'' model for formation of close-in giant planets is that gas giants may be originally formed at several AU's beyond ice lines and migrate to the vicinity of the star.
The most referred migration mechanism is type II migration.
Here, we have investigated another channel to move planets to the stellar vicinity, ``scattering'' model (Rasio \& Ford 1996), which is a coupled process of planet-planet scattering and tidal circularization. 
If orbital eccentricity is pumped up to values close to unity, because of the proximity of the pericenter the eccentricity and the semimajor axis of the planet are damped by the tidal effect from the star, almost keeping the pericenter distance to form a close-in planet with relatively small eccentricity.
We newly found that Kozai mechanism is also one of key processes in the model.  

We have carried out orbital integration of three planets of Jupiter-mass, including the tidal damping in the integration.
We follow the simulation setup by Marzari and Weidenschilling (2002).
We found that in $\sim$30\% runs, close-in planets are formed, which is much higher probability than suggested by previous studies (Weidenschilling and Marzari 1997; Marzari and Weidenschilling 2002; Chatterjee et al. 2007), because in many cases, the circularization occurs during three-planet scattering, which was not monitored by previous studies. 
The three-planet scattering is usually terminated by ejection of one planet and the system enters a stable state.
Previous studies were concerned only with the stable state.

Since the tidal damping timescale is usually longer than that of orbital change in the chaotic stage of three-planet interaction, the tidal circularization does not occur during the chaotic stage, even if the pericenter distance happens to be small enough.
But, we have found that when one planet is scattered inward, it often becomes separated from other outer planets.
As long as the tidal dissipation is negligible, the isolation is only tentative, but its duration is long enough for the tidal circularization if the pericenter distance is small enough 
($\la 0.02$--0.04 AU for Jupiter-mass planets).
We also found out that the isolated planet usually enters Kozai state. 
Even if the eccentricity of the planet is not excited enough by the close scattering that injected the planet to an inner orbit, the eccentricity can secularly increase to values close to unity during Kozai cycle, in particular, when the planet has an inclined orbit. 
Although the probability that the eccentricity takes values between 0.98 and 1.0 is quite low just after the scattering, that probability is significantly enhanced to 30\% level during the Kozai cycle.
Even if the Kozai cycle has relatively large angular momentum and the eccentricity does not take such high value, the outer planets eventually destroy the quasi-stable state and have the inner planet enter another quasi-stable state, in other words, another Kozai state.
The new Kozai state can have relatively small angular momentum and raise the eccentricity high enough for the tidal circularization. 
The repeated Kozai mechanism enhances the probability for the tidal circularization to occur.
Thereby, we have found much higher probability of formation of close-in planets than previously expected.
Therefore, the scattering would contribute to formation of close-in planets, although a main channel may still be through type II migration.

\begin{figure}\epsscale{0.8}
\plotone{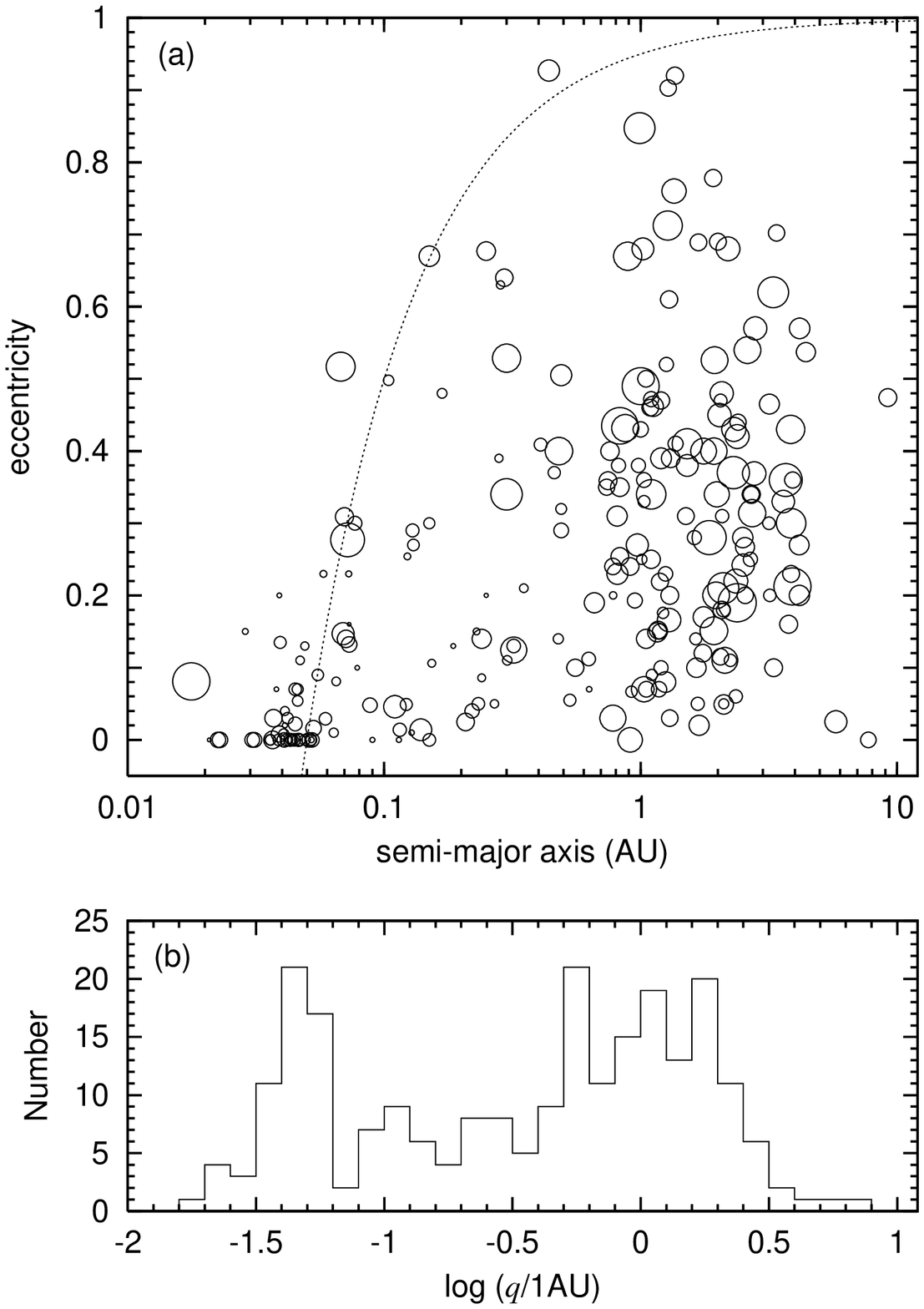}
\caption{(a) Distribution of orbital eccentricity ($e$)
and semimajor axis ($a$) of observed extrasolar planets. 
Circle sizes are proportional to $(m\sin i)^{1/3}$. Dotted line shows 
pericenter distance $q=a(1-e)=0.05$ AU. (b) Histogram of 
pericenter distance. 
\label{fig:exso}}
\end{figure}

\begin{figure}\epsscale{0.8}
\plotone{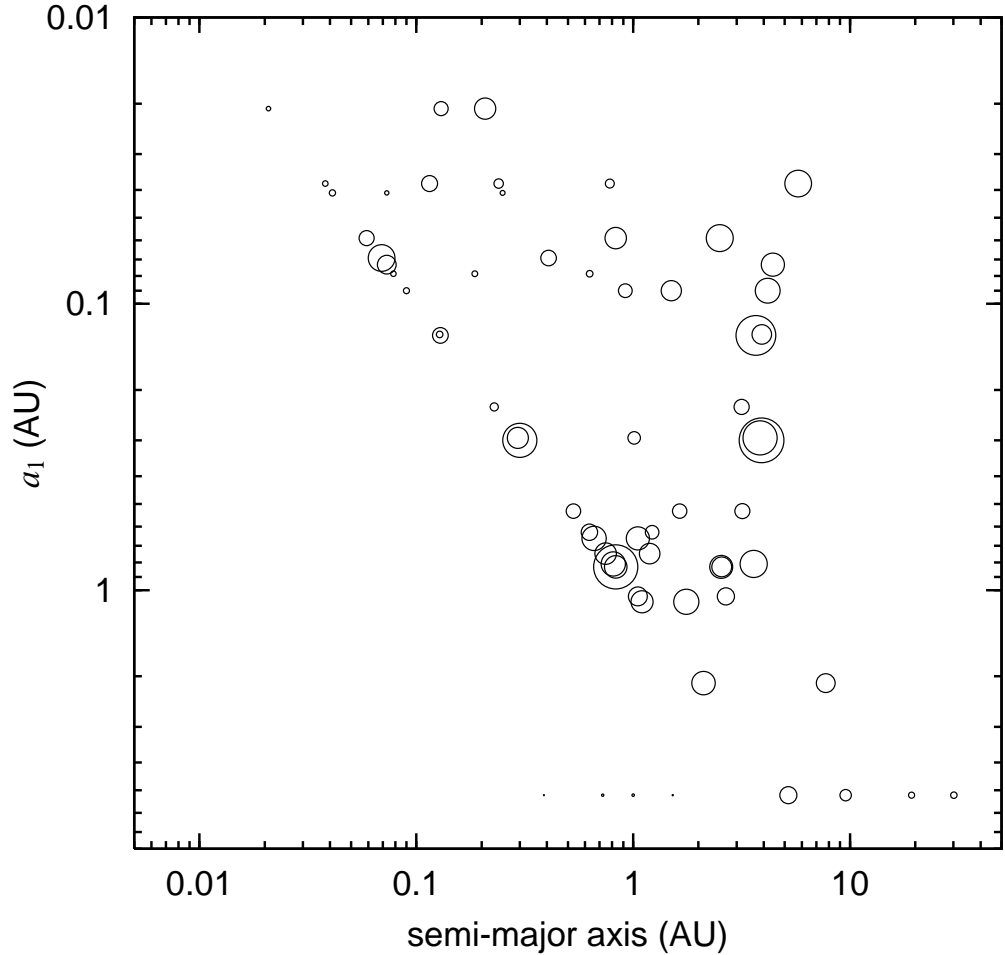}
\caption{Semi-major axes of extrasolar planets in multiple planetary 
systems. The vertical axis is the semi-major axis ($a_1$) of an
innermost planet in each system.
In the bottom, the solar system is also shown with $a_1=a_{\rm J}$. 
Circle sizes of extrasolar planets are proportional to 
$(m\sin i)^{1/3}$. 
\label{fig:multi}}
\end{figure}

\begin{figure}\epsscale{0.8}
\plotone{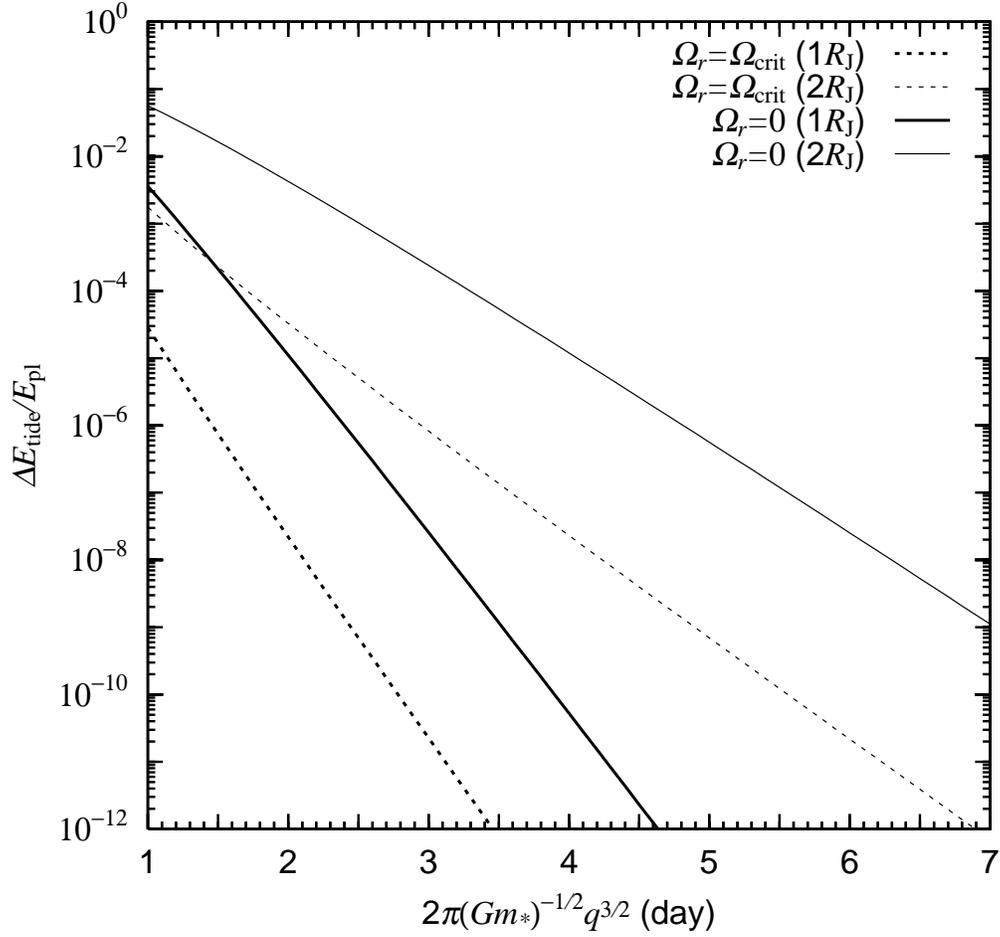}
\caption{The energy change $\Delta E_{\rm tide}$ during
one pericenter passage as a function of orbital period.
$E_{\rm pl}$ is orbital energy.  
A solar-mass host star and a Jupiter-mass planet are assumed.
For larger planetary mass, $\Delta E_{\rm tide}$ is larger.
Solid lines show 
the change for non-rotating planets. Dotted lines are calculated for 
planets with critical rotation which leads to $\Delta L_{\rm tide}=0$. 
Thick lines correspond to  $1 R_{\rm J}$ radius planets. Thin lines are 
for $2 R_{\rm J}$ planets. 
\label{fig:deleng}}
\end{figure}

\begin{figure}\epsscale{0.8}
\plotone{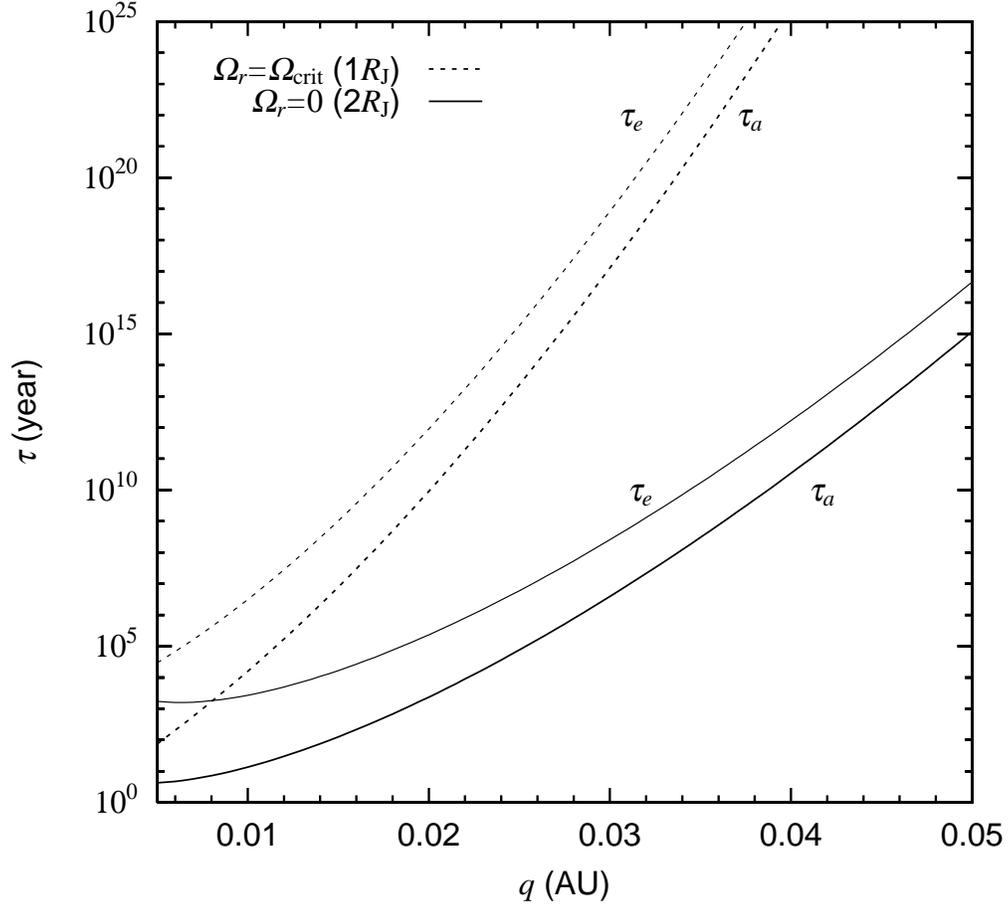}
\caption{Tidal circularization timescales.
It is given as a function of
pericenter distance $q$.
The formulae are given by  
equations (\ref{eq:a_damp_timescale})
and (\ref{eq:e_damp_timescale}) with 
equations (\ref{eq:lnorot}) and (\ref{eq:enorot}) for
non-rotating planets or
equations (\ref{eq:lcrit}) and (\ref{eq:ecrit}) for
planets with the critical rotation, following Ivanov \& Papaloizou (2004).
Solid lines are for 
non-rotating planets with $R=2R_{\rm J}$. 
Dotted lines are for 1$R_{\rm J}$ planets 
rotating with the critical rotation. Thick lines show 
$\tau_a$ and thin lines show $\tau_e$. 
The semi-major axis $a=$2 AU is 
used for the estimation.
A solar-mass host star and a Jupiter-mass planet are assumed.
For larger planetary mass, the timescales are much shorter.
\label{fig:timescale}}
\end{figure}

\begin{figure}\epsscale{0.8}
\plotone{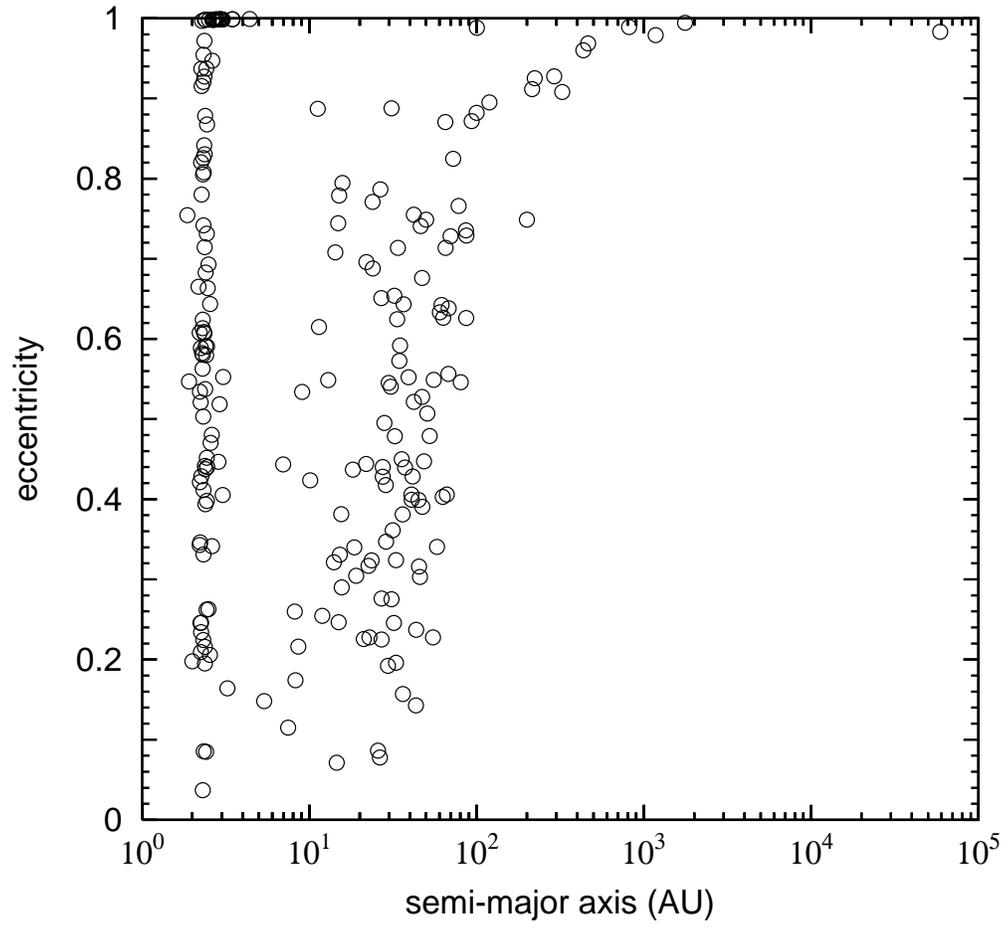}
\caption{Final distribution of $a$ and $e$ in the case of Set N (without 
tide). 
\label{fig:nonea}}
\end{figure}

\begin{figure}\epsscale{1.0}
\plotone{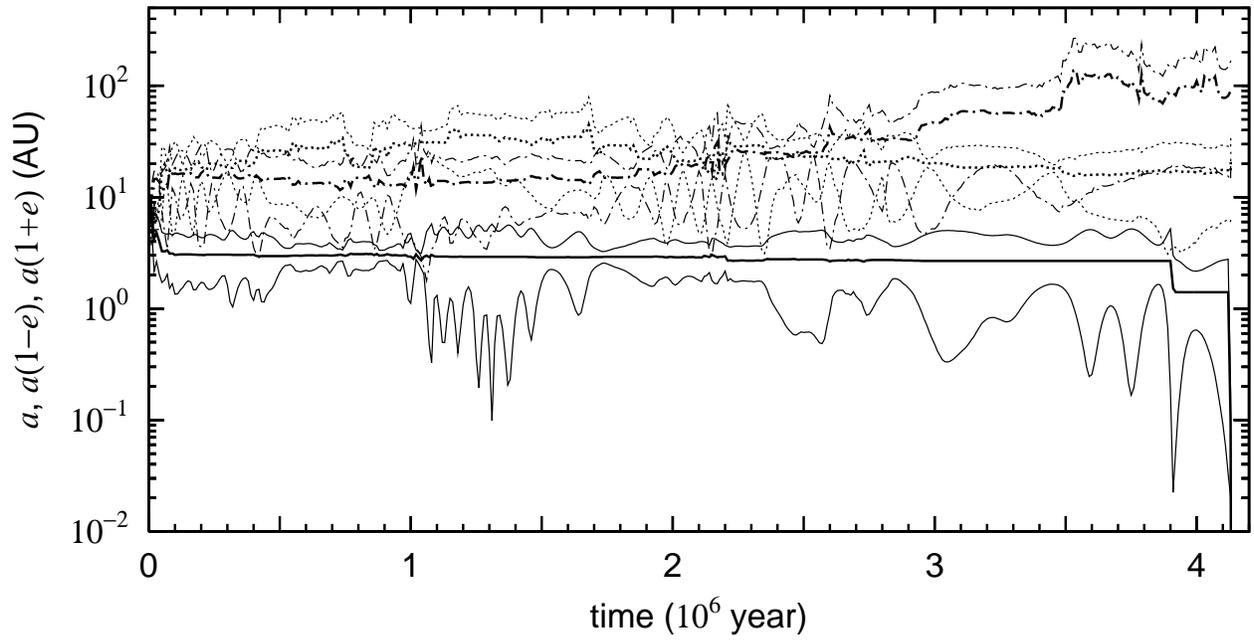}
\caption{Typical evolution of the semi-major axes ($a$) of three planets. 
Thin lines show evolution of pericenters ($a(1-e)$) 
and apocenters ($a(1+e)$). The planet indicated 
by solid line is circularized at $4.13\times 10^6$ years. 
\label{fig:ex1}}
\end{figure}

\begin{figure}\epsscale{1.0}
\plotone{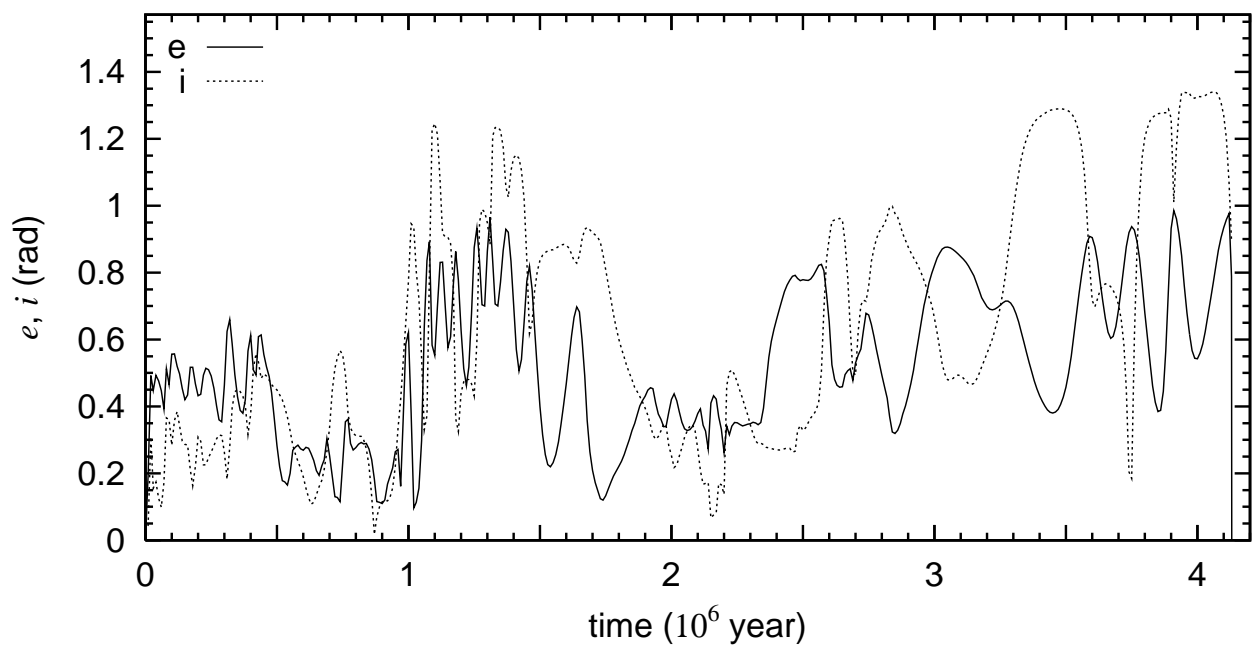}
\caption{Evolution of eccentricity $e$ and inclination $i$ 
of the circularized 
planet shown in Figure \ref{fig:ex1}. Solid and dotted lines indicate 
$e$ and $i$, respectively.  
\label{fig:ex1ei}}
\end{figure}

\begin{figure}\epsscale{1.0}
\plotone{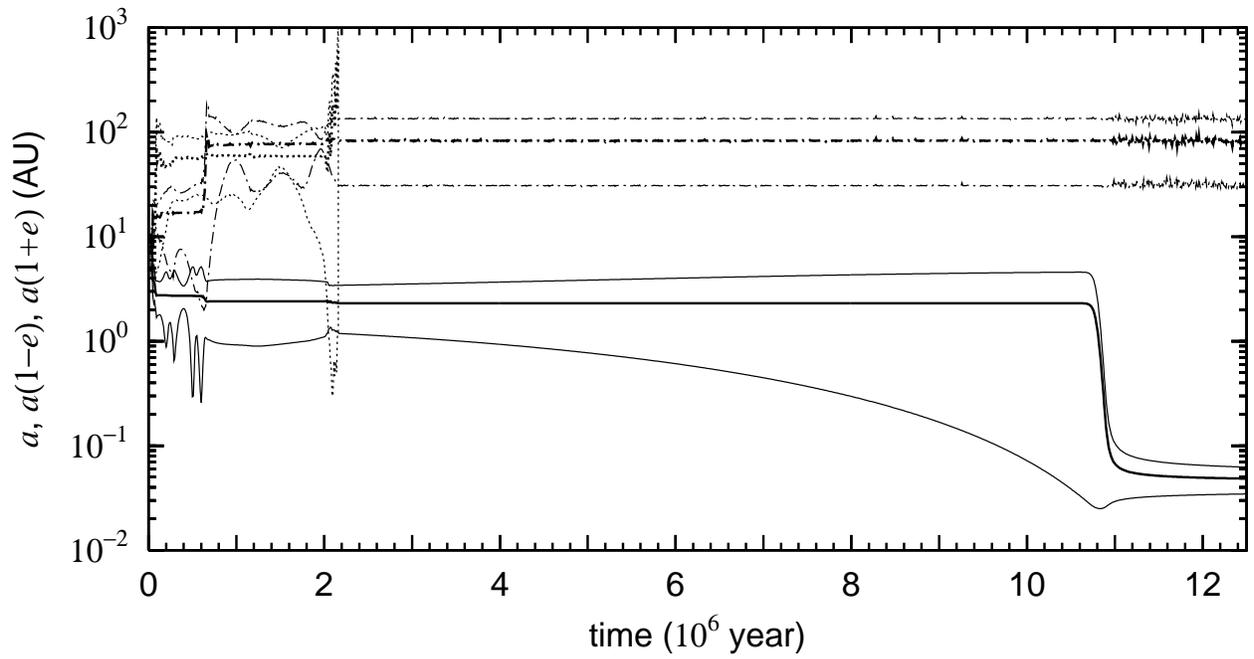}
\caption{Evolution of the semi-major axes of three planets. The meaning 
of the lines is the same as Figure \ref{fig:ex1}. One planet shown by 
dotted line is ejected from the system at 2.2 $\times 10^6$ years. The planet 
indicated by solid line is circularized at $\simeq 1.1\times 10^7$ years. 
\label{fig:ex2}}
\end{figure}

\begin{figure}\epsscale{0.8}
\plotone{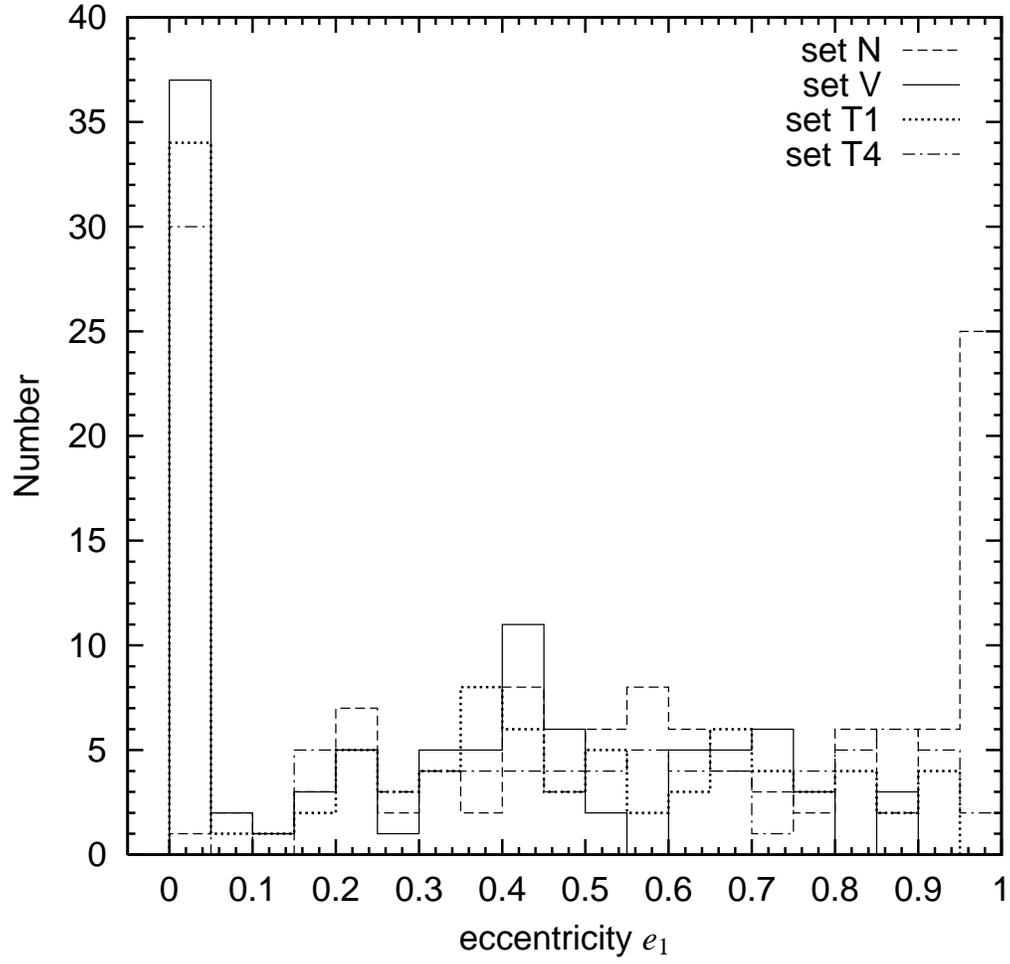}
\caption{Eccentricity histogram of the planets that were scattered into 
inner orbits. Long-dashed, solid, dotted, and dash-dotted lines 
correspond to Sets N, V, T1 and T4, respectively. 
\label{fig:histe}}
\end{figure}

\begin{figure}\epsscale{0.8}
\plotone{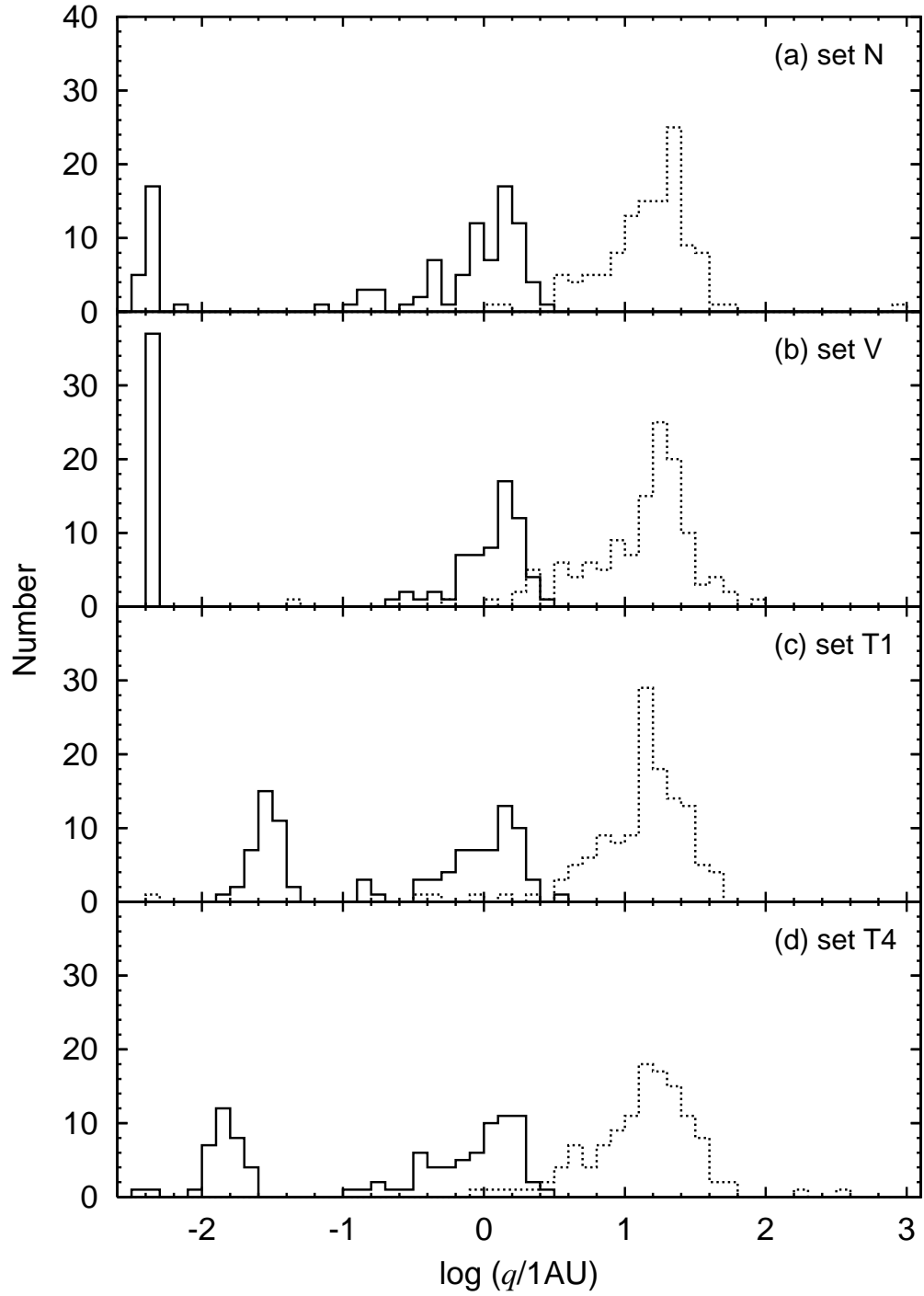}
\caption{Histogram of pericenter distance. 
Solid lines represent inner planets 
and dotted lines represent outer ones in the end of simulations. 
\label{fig:histq}}
\end{figure}

\begin{figure}\epsscale{0.67}
\plotone{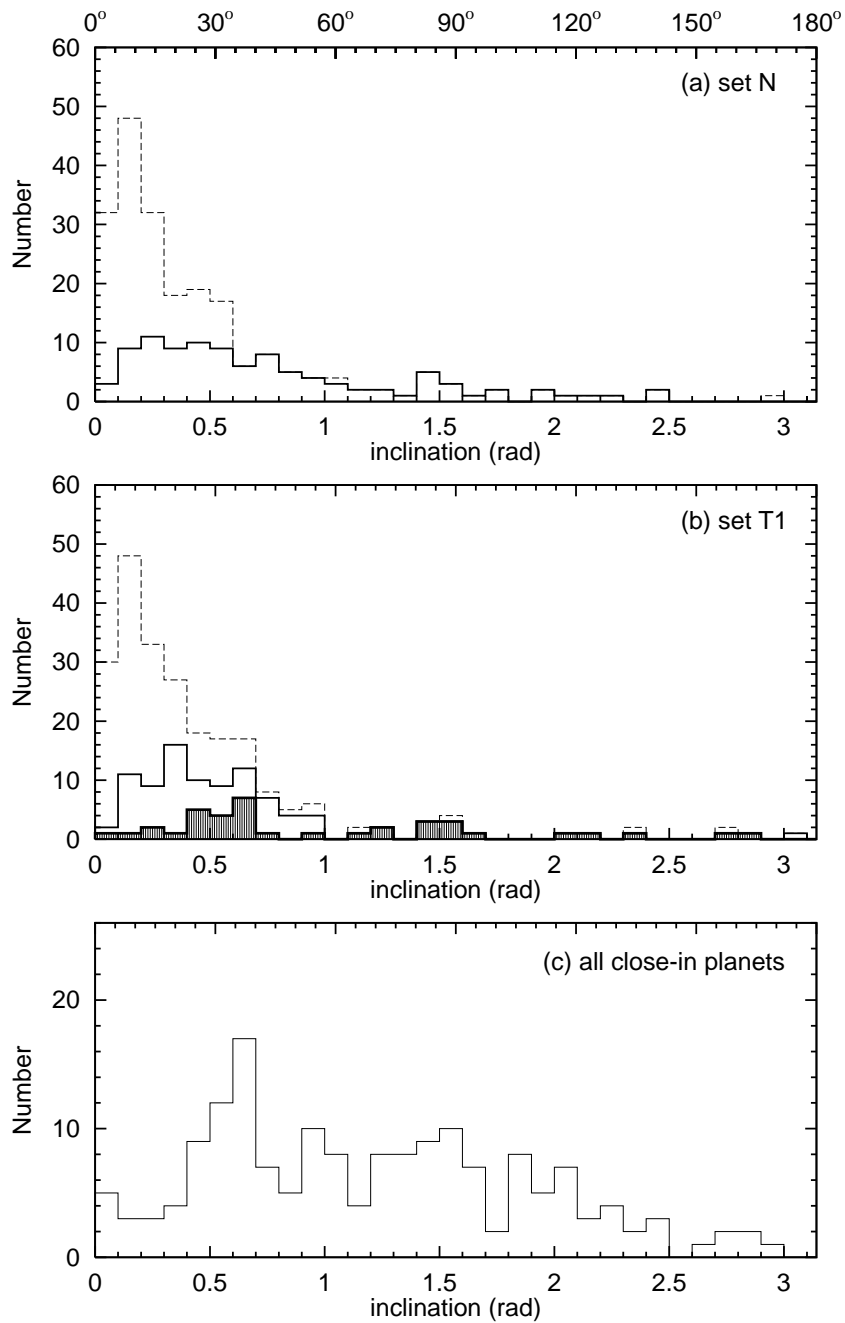}
\caption{Histogram of inclination of planets. 
(a) In the case without tidal force (Set N).
Solid line shows the innermost planets. Dashed lines are all surviving planets. 
The planets hit the star are counted as surviving planets. 
(b) In the case with tidal force (Set T1). 
The inclination of close-in planets is shown in the shaded region. 
(c) The inclination of formed close-in planets in all simulations (Set V and Sets T1-T4). 
\label{fig:histi}}
\end{figure}

\begin{figure}\epsscale{1.0}
\plotone{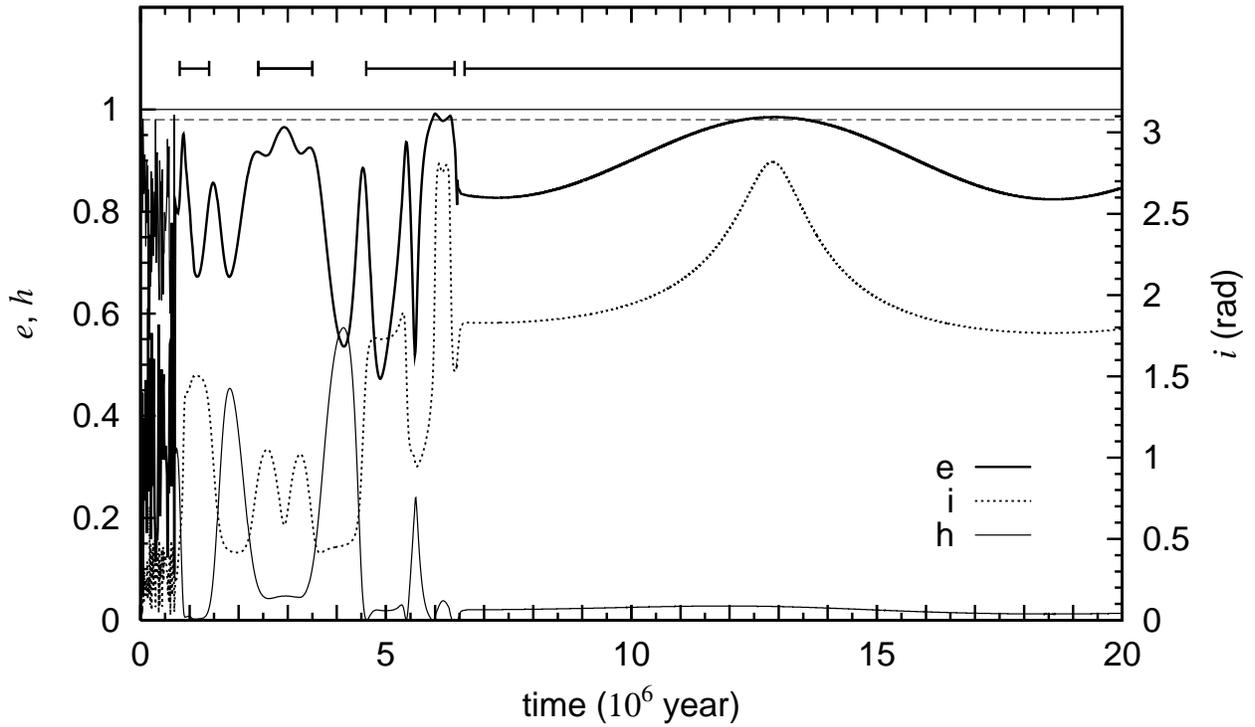}
\caption{An example of evolution of $e$, $i$, and $h$ of the inner planet. 
The evolution of the inclination is shown in secondary axis. Dashed line 
shows $e=0.98$. The bars in the figure show the period that the 
planet has small $h$ value, i.e., it is in Kozai state.
\label{fig:nonlong}}
\end{figure}

\begin{figure}\epsscale{1.0}
\plotone{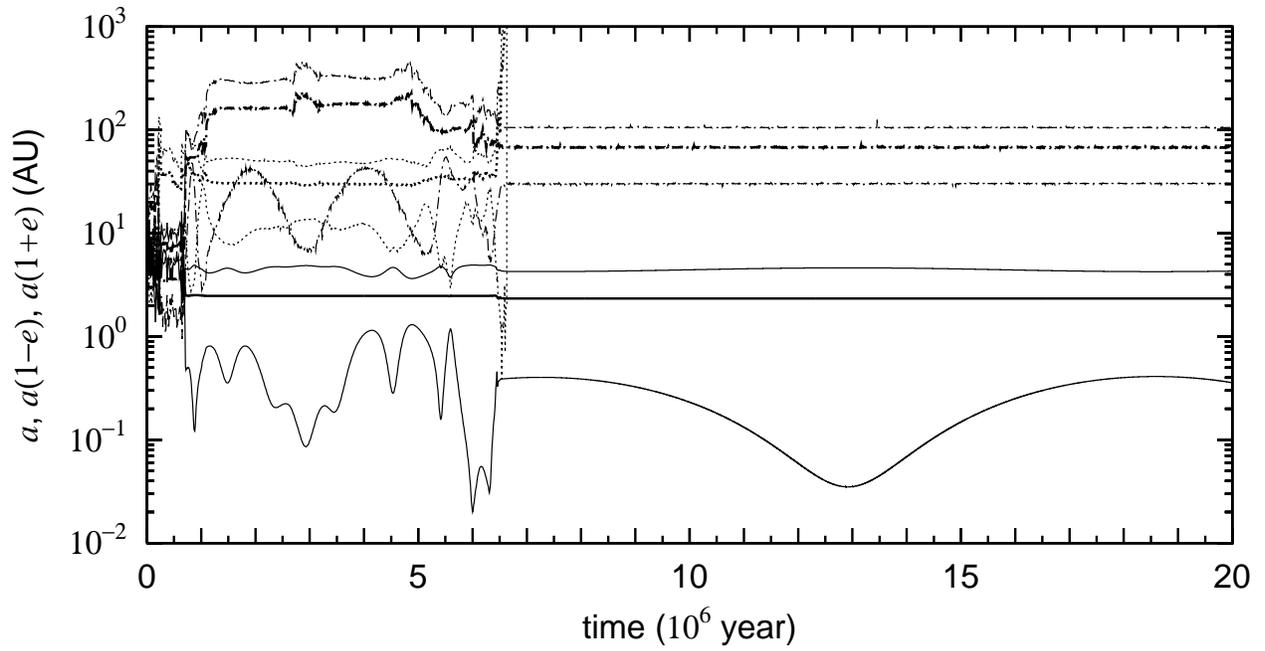}
\caption{Semi-major axis, pericenter, and apocenter evolution in the 
system corresponding to Figure \ref{fig:nonlong}.
\label{fig:nonlonga}}
\end{figure}

\begin{figure}\epsscale{1.0}
\plotone{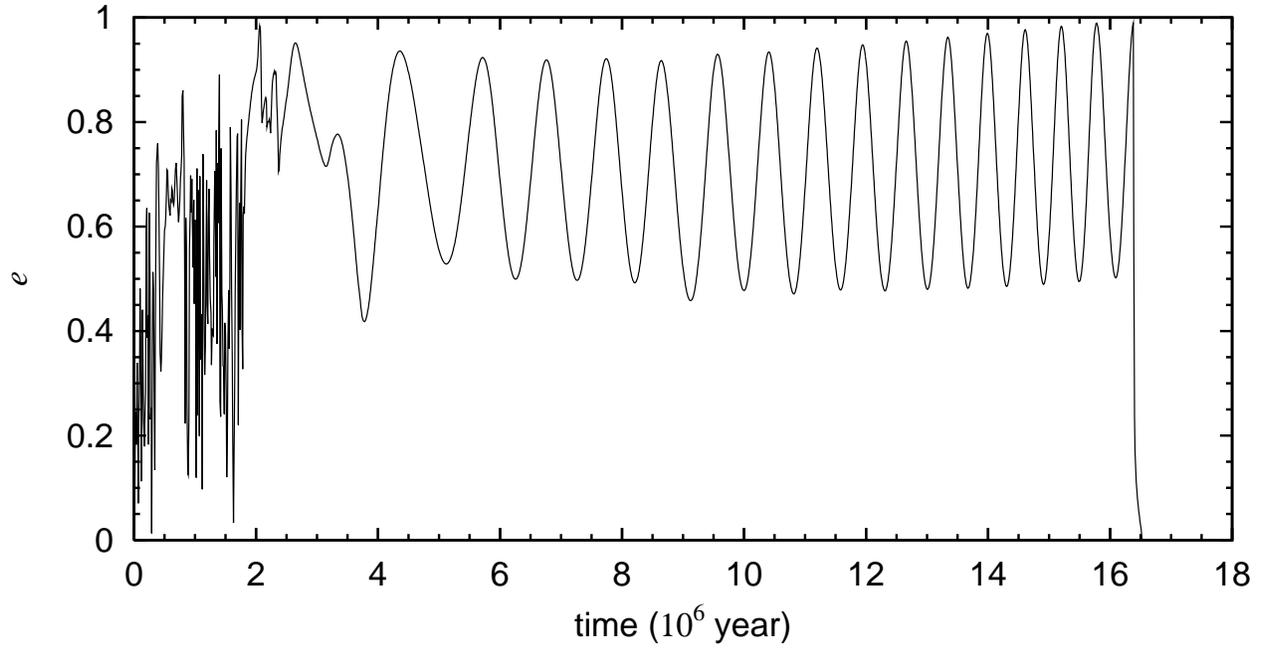}
\caption{An example of evolution of eccentricity of the inner planet.
\label{fig:oscie}}
\end{figure}

\begin{figure}\epsscale{0.59}
\plotone{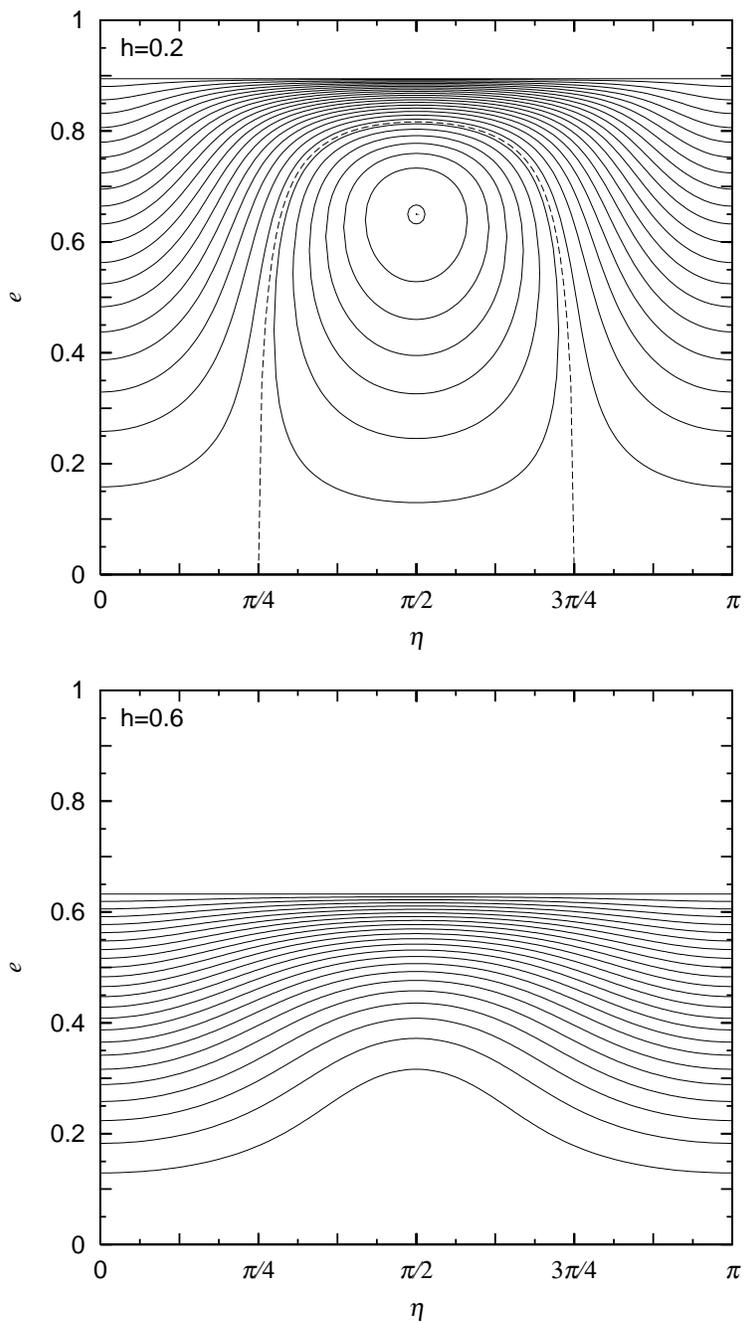}
\caption{Hamiltonian map for an innermost planet with $h=0.2$ and $h=0.6$. 
The contours in upper panel
are drawn at every $\Delta C=0.5$ from $C = -4.02$ to 8.8
(inner ones to outer ones) for $h=0.2$.
$C$ takes the minimum ($\simeq -4$) at libration center 
and the maximum at the top contour ($=8.8$).
In lower panel, contours are drawn at every
$\Delta C=0.2$  from $C = 1.6$ to 6.4 (lower ones to upper ones)
for $h=0.6$. 
The dotted line shows a boundary between libration 
and circulation. 
}
\label{fig:hamap}
\end{figure}

\begin{figure}\epsscale{0.9}
\plotone{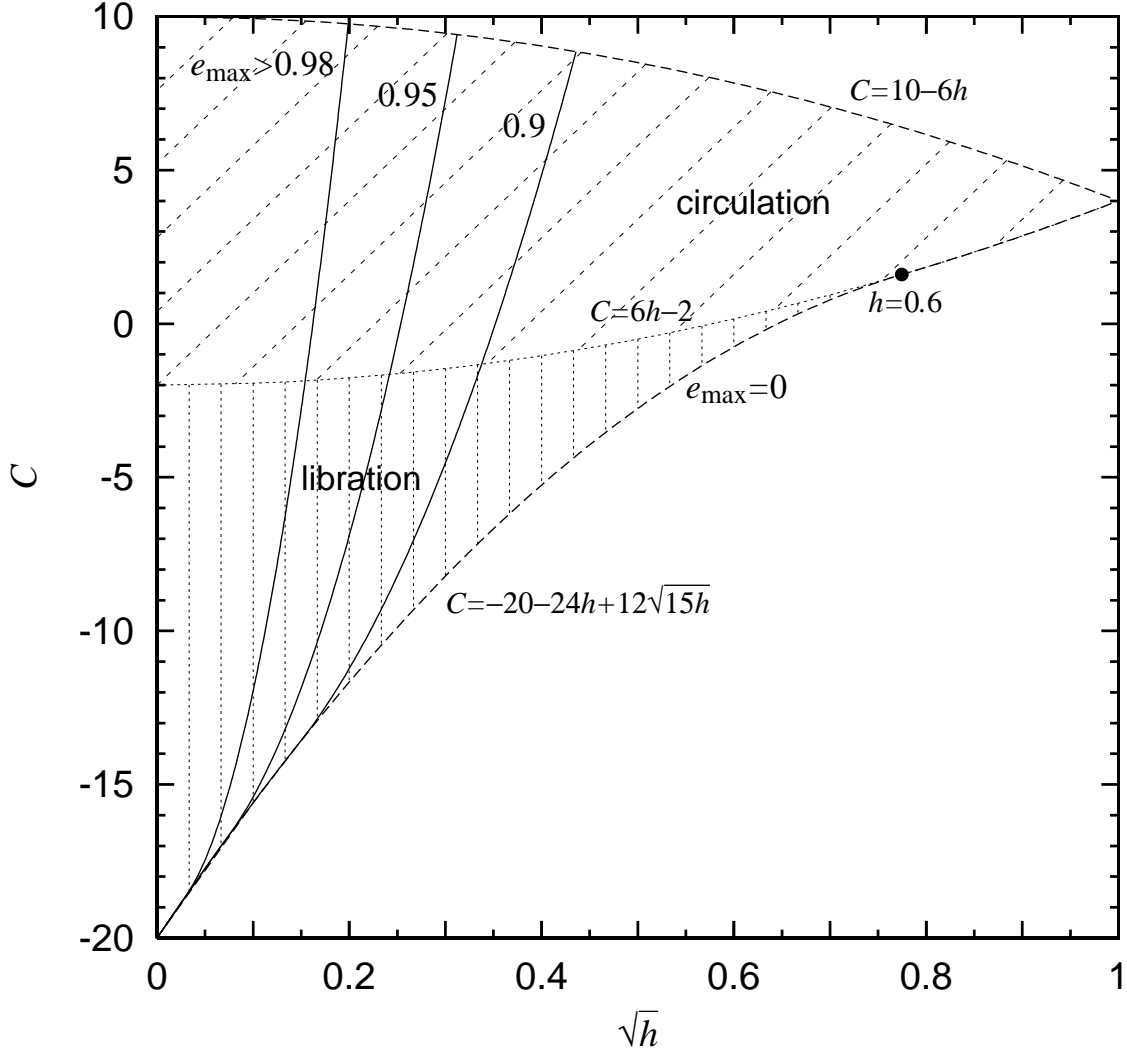}
\caption{The range of $C$ and $\sqrt{h}$ for given $e_{\rm max}$. 
In $h \le 0.6$, the 
minimum value of $C$ is $C=-20-24h+12\sqrt{15h}$. In $h \ge 0.6$, it 
is given by $C=6h-2$. 
In shaded regions (denoted by 'libration' and 'circulation'), the planet is in an elliptical orbit. 
In the region denoted by libration, the mutual angle of pericenters ($|\eta|$) is restricted between $\pi/4$ to $3\pi/4$. 
In the circulation region, $|\eta|$ circulates from 0 to $\pi$.
\label{fig:chrange}}
\end{figure}

Non-negligible number of close-in planets without nearby secondary planets but with 
eccentricities $\ga 0.3$ has been discovered.
It is not easy for type II migration model to account for the high eccentricities, 
but the scattering model could account for them.
As shown in sections 3.3 and 3.4, the tidal circularization can slow down
by increase in $q$ before full circularization, leaving moderate eccentricities.
The discovered close-in eccentric planets are generally 
massive (having masses larger than Jupiter-mass)
among the close-in planets (Figure \ref{fig:exso}a), so their circularization timescales 
are much shorter than those shown in Figure \ref{fig:timescale}.
Therefore, the results in our Sets T could be consistent with these discovered planets.

Our results also suggest that the close-in planets have a broadly spread inclination distribution, including retrograde rotation.
More observations of the Rossiter-McLaughlin effect for the transiting close-in planets,
in particular for those with relatively high eccentricity such as HD17156b,
will impose constraints on the contribution of the scattering model to close-in planets. 

The tidally dissipated energy is $\sim G M_{\ast} m/2a$ in total, where $a$ is final semi-major axis of the circularized close-in planet and $m$ is its planet mass.
Since this energy is much larger than binding energy of the planet, it could be heat source for inflated close-in planets such as HD209458b, OGLE-TR-10, OGLE-TR-56, HAT-P-1b or WASP-1b, if the tidal circularization occurred relatively recently (orbital crossing followed by the circularization can start after long time after the formation of giant planets).

Note that the tidal circularization model we used may not be sufficiently relevant. We did many assumptions for the usage of equations (\ref{eq:ang}) and (\ref{eq:eng}). 
The model is only valid for fully convective planets in highly eccentric orbits, since it consider only $\ell=2$ f-mode and use impulse approximation.
We apply the model even after the orbits are significantly circularized, which may overestimate effect of tidal dissipation.  
In the multiple encounters to the star, the energy and angular momentum can either increase or decrease (e.g., Press \& Teukolsky 1997; Mardling 1995a, 1995b), but we include the tidal effect as it always gives decreasing effect. 
Because of these simplifications, our simulation may be a limiting case that the tidal force works most effectively, although inclusion of g-modes might strengthen the tidal effect.
Our purpose of this paper is not a study of the detailed tidal damping process but to show another available channel to the close-in planets. 
We also need to take into account the relativity precession and J2 potential of the central star which prevent the Kozai mechanism, as well as long-term tidal disruption process (e.g., Gu, Lin, \& Bodenheimer 2003).
These issues are left for further studies.

\acknowledgments 
We would like to thank E. Kokubo, A. Morbidelli, and an anonymous referee for their helpful suggestions. 
This work is supported by MEXT KAKENHI(18740281) Grant-in-Aid for Young 
Scientists (B), Grant-in-Aid for Scientific Research on Priority Areas
(MEXT-16077202), and MEXT's program "Promotion of Environmental 
Improvement for Independence of Young Researchers" under the Special 
Coordination Funds for Promoting Science and Technology.

\clearpage

\clearpage

\clearpage

\clearpage

\clearpage

\clearpage

\clearpage

\clearpage

\clearpage

\clearpage

\clearpage

\clearpage

\clearpage

\clearpage

\clearpage

\clearpage

\clearpage
\begin{deluxetable}{ccccc}
\tablecolumns{5}
\tablewidth{0pc}
\tablecaption{Characteristics of simulations}
\tablehead{
\colhead{Set} & \colhead{$\Omega_r$}& \colhead{$R/R_{\rm J}$}& 
\colhead{comments}& \colhead{Close-in planets}
}
\startdata
Set N \ & - & - & No tide & - \\
Set V \ & 0 & 2 & $\Delta v^2=2\Delta E_{\rm tide}$& 37\% \\
Set T1 & 0 & 2 & equation (\ref{eq:lnorot})&  38\% \\
Set T2 & 0 & 1 & equation (\ref{eq:lnorot})& 29\%  \\
Set T3 & $\Omega_{\rm crit}$ & 2 & equation (\ref{eq:lcrit})& 33\% 
\\
Set T4 & $\Omega_{\rm crit}$ & 1 & equation (\ref{eq:lcrit})& 32\% 
\\
\enddata \label{tab:set}
\end{deluxetable}

\end{document}